\documentclass{pasj00}
\usepackage{times}
\usepackage{mathptm}

\begin{document}
\SetRunningHead{H. Nakanishi, Y. Sofue, and J. Koda}{Virgo High-Resolution CO Survey :V. CER in NGC 4569}
\Received{2002/10/23}
\Accepted{2005/10/23}
\Published{2005/12/25}
\title{Virgo High-Resolution CO Survey :\\
V. Circumnuclear Elliptical Ring in NGC 4569} 

\author{Hiroyuki \textsc{Nakanishi},\altaffilmark{1,2} Yoshiaki
\textsc{Sofue},\altaffilmark{1} and Jin \textsc{Koda}\altaffilmark{1,3,4}}%
\email{hnakanis@nro.nao.ac.jp}

\altaffiltext{1}{Institute of Astronomy, The University of Tokyo, 2-21-1 Osawa, Mitaka,Tokyo 181-0015}
\altaffiltext{2}{Nobeyama Radio Observatory, Minamimaki, Minamisaku, Nagano 384-1305}
\altaffiltext{3}{National Astronomical Observatory of Japan, 2-21-1
Osawa, Mitaka,Tokyo 181-8588}
\altaffiltext{4}{California Institute of Technology, MC 105-24, Pasadena, CA 91125, USA.}
\KeyWords{galaxies: clusters: individual (Virgo) --- galaxies: individual (NGC 4569, M 90) --- galaxies: ISM --- galaxies: kinematics and dynamics --- galaxies: spiral } 

\maketitle
\begin{abstract}
We present high-resolution ($1\farcs8$ -- $4\farcs5$) CO data of the Virgo spiral galaxy NGC 4569, obtained using the Nobeyama Millimeter Array. We found that the molecular gas is highly concentrated in the circumnuclear region with two off-center peaks. A CO image with the highest angular resolution of $2\farcs0 \times  1\farcs8$ shows that six blobs likely form a circumnuclear elliptical ring (CER) with a semi-major axis radius of 720 pc. 
The CER shows a strongly twisted velocity field, and the position--velocity (PV) diagram shows significant forbidden velocity components. These kinetic features are understood as being non-circular motion due to a bar-potential. We found that the CER coincides with the H$\alpha$ bright central core and that the mass ratio of the molecular gas to the dynamical mass is about 18\%. These results support a gaseous inflow scenario induced by a weak bar potential and self-gravity of the gas.
\end{abstract}

\section{Introduction}
NGC 4569 is a member of the Virgo cluster galaxies and its nucleus is classified as a starburst-dominated LINER \citep{alo00}. 
The mechanism for causing nuclear activity is still unveiled. A CO line observation is useful to explore the physical conditions of the central regions of spiral galaxies to understand this issue, because the molecular gas is a dominant component among neutral gases in the central region. 

\citet{sak99} observed this galaxy using the Nobeyama Millimeter Array (NMA) at the Nobeyama Radio Observatory (NRO) in the $^{12}$CO($J=1$--$0$) line, and obtained a resolution of $3\farcs9 \times 3\farcs6$. They showed the gas concentration in the central region, two significant peaks on the central concentration, and strong non-circular motions. \citet{hel03} obtained $^{12}$CO($J=1$--$0$) line data with a resolution of $6\farcs6 \times 5\farcs5$ in the BIMA SONG project. They found extended CO emission up to about $1\arcmin$ radius.

In order to investigate the more detail kinematics in the central region of spiral galaxies, 
we achieved the Virgo high-resolution CO-line survey (ViCS) during a long-term project at the NMA from 1999 December to 2000 March to obtain the highest resolution CO data (Sofue et al., 2003, Paper I hereafter ). We presented CO data of NGC 4569 with low-resolution ($4\farcs5 \times 3\farcs1$) in Paper I.  
In this paper as Part V of the series of the ViCS, we present the highest resolution ($2\farcs0\times 1\farcs8$) CO data, and discuss the distribution of the interstellar matter (ISM) in the central region and the relationship to the nuclear activity. 

\begin{figure}
  \begin{center}
    \FigureFile(80mm,80mm){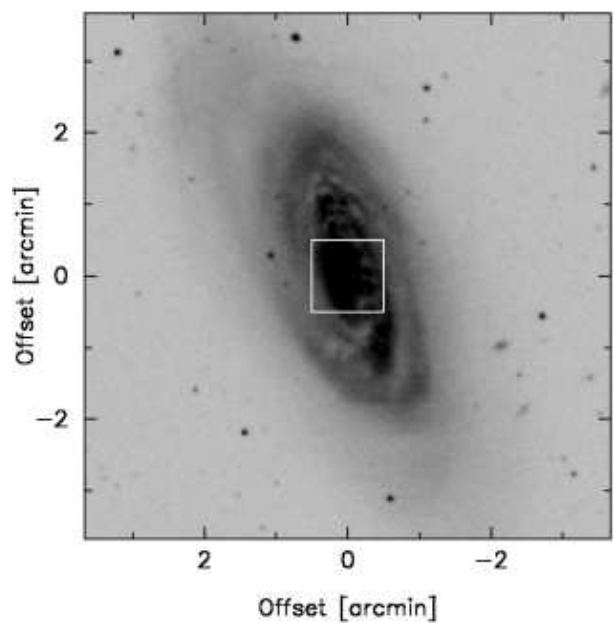}
  \end{center}
  \caption{STScI Digitized Sky Survey (DSS) $B$-band image of NGC 4569. The central square indicates regions of CO maps ($1\arcmin \times 1\arcmin$), which are shown in figure \ref{MOM0}.  \label{dss}}
\end{figure}

An optical $B$-band image of this galaxy is shown in figure \ref{dss}, which was taken from STScI Digitized Sky Survey (DSS). The properties of this galaxy are summarized in table \ref{prop-n4569}.
NGC 4569 is assumed to be at a distance of $16.1$ Mpc, which is the distance of the Virgo cluster member NGC 4321 \citep{fer96}. The angular size of $1\arcsec$ corresponds to 78 pc. {A position angle (P.A.) of $18\arcdeg$ and an inclination of $63\arcdeg$ are adopted \citep{nak04}. }

\vskip 3mm
\centerline{--- Figure \ref{dss} ---}
\vskip 3mm

\vskip 3mm
\centerline{--- Table \ref{prop-n4569} ---}
\vskip 3mm

\section{Observation and Reduction}
We carried out interferometer observations of the central region of NGC 4569 in the $^{12}$CO ($J=1$--$0$) line using the NMA in the AB, C, and D configurations. {D, AB, and C-configuration observations were made in 1999 December, 2000 February, and 2000 March, respectively, as a part of the ViCS project. 
}

The observation time was typically 7 -- 8 hours per each configuration. 
The pointing center was at RA(J2000) $= 12^{\rm{h}}36^{\rm{m}}49\fs82$, Dec(J2000) $= +13\arcdeg 09\arcmin 45\farcs8$, adopted from \citet{sak99}. 
For flux and phase calibrations, we observed the nearby radio point source 3C 273 every 20 minutes {with an integration time of 3 minutes. We calibrated the flux against NRAO 530, 1749+096, and 0507+179. The fluxes of the gain calibrator, 3C 273, were 11.2 Jy in the D-configuration observation, 10.4 Jy in the AB-configuration observation, and 11.2 Jy in the C-configuration observation. The obtained fluxes contained errors of $\sim 20 \%$. }
We used a spectro-correlator system, Ultra Wide Band Correlator (UWBC) \citep{oku00}, in a narrow-band mode, which had 256 channels and the total bandwidth of 512 MHz. One channel corresponded to 5.2 km s$^{-1}$ at the observing frequency. {Table \ref{obs-parm} gives the observational parameters.}
 
\vskip 3mm
\centerline{--- Table \ref{obs-parm} ---}
\vskip 3mm

{The raw data were calibrated using the reduction software UVPROC-II, developed at NRO. A baseline correction, a system temperature ($T_{\rm sys}$) correction, a bandpass calibration, and a flux calibration were applied. The calibrated UV data were Fourier-transformed using the National Radio Astronomy Observatory (NRAO) Astronomical Image Processing System (AIPS) to derive CO images. }
We adopted the CLEAN method and applied two different weighting functions for the UV data, to produce two maps with different resolutions and sensitivities (see table \ref{parm-map}).

We first obtained a CO image using a natural weighting function. The data were averaged in 2 bins (10.4 km s$^{-1}$) of the original channels, and the channel increment was 2 (10.4 km s$^{-1}$), yielding a 128-channel data cube. A pixel size of $0\farcs25$ was adopted for the mapping. The synthesized beam size (full width half maximum; FWHM) was $4\farcs5\times 3\farcs1$, and the typical r.m.s. noise on a channel map was 23 mJy per beam. 

Second, we obtained a high-resolution map using a uniform weighing function. The data were averaged in 4 bins, giving a velocity resolution of 20.8 km s$^{-1}$. The channel increment was fixed at 2 (10.4 km s$^{-1}$). The synthesized beam size (FWHM) was $2\farcs0\times 1\farcs8$, which was the highest resolution among ever obtained data. The typical r.m.s. noise was 34 mJy per beam. The parameters of the maps are given in table \ref{parm-map}.

\vskip 3mm
\centerline{--- Table \ref{parm-map} ---}
\vskip 3mm

\section{Results}
{In this section, we present the obtained spectra, channel maps, CO intensity maps, a velocity field, Position--Velocity (PV) diagrams, and comparisons with the optical broadband, H$\alpha$, and H {\sc i} images. The highest resolution CO map first { shows that there may exist an elliptical ring} in the circumnuclear region (subsection \ref{sec-map}), which has strong non-circular motions due to a bar-potential (subsections \ref{sec-vel} and \ref{sec-pv}). 
}

\subsection{Spectra}
\begin{figure*}
  \begin{center}
    \FigureFile(160mm,80mm){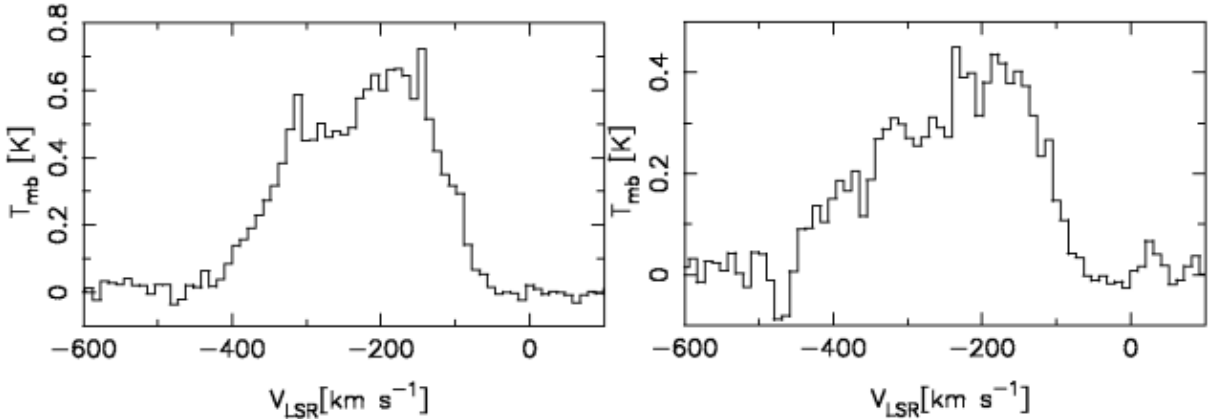}
  \end{center}
  \caption{CO spectra at the observing centers of the low- and high-resolution maps, corrected for primary beam attenuation and convoluted with a Gaussian profile (FWHM $= 16\arcsec$). Left: Spectrum of the low-resolution map. Right: Spectrum of the high-resolution map. Because of large missing flux in the high-resolution map, the right spectrum is weaker than the left one. \label{spectra}}
\end{figure*}

{Figure \ref{spectra} shows CO line spectra at the center of NGC 4569. The top and bottom panels present spectra of the low- and high-resolution maps, respectively. In order to compare them with a single dish flux from the NRO 45 m telescope (FWHM $= 16\arcsec$), the data were corrected for primary beam attenuation, convoluted with a Gaussian profile (FWHM=$16\arcsec$), and sampled at the pointing center. 
The integrated intensities of the low- and high-resolution maps were 149 K km s$^{-1}$ and 96 K km s$^{-1}$, respectively. The Nobeyama 45-m telescope observation of NGC 4569 (Nishiyama, Nakai, 2001) showed an integrated intensity of 134 K km s$^{-1}$.

The fractions of the single-dish flux recovered by the aperture synthesis observations were 111\% and 71\% in the cases of the low- and high-resolution maps, respectively. The spectrum at the bottom of figure \ref{spectra} is weaker than that at the top, because the missing flux is larger in the high-resolution map than in the low-resolution map.  
}

\vskip 3mm
\centerline{--- Figure \ref{spectra} ---}
\vskip 3mm

\subsection{Channel Maps}
\begin{figure*}
  \begin{center}
    \FigureFile(160mm,240mm){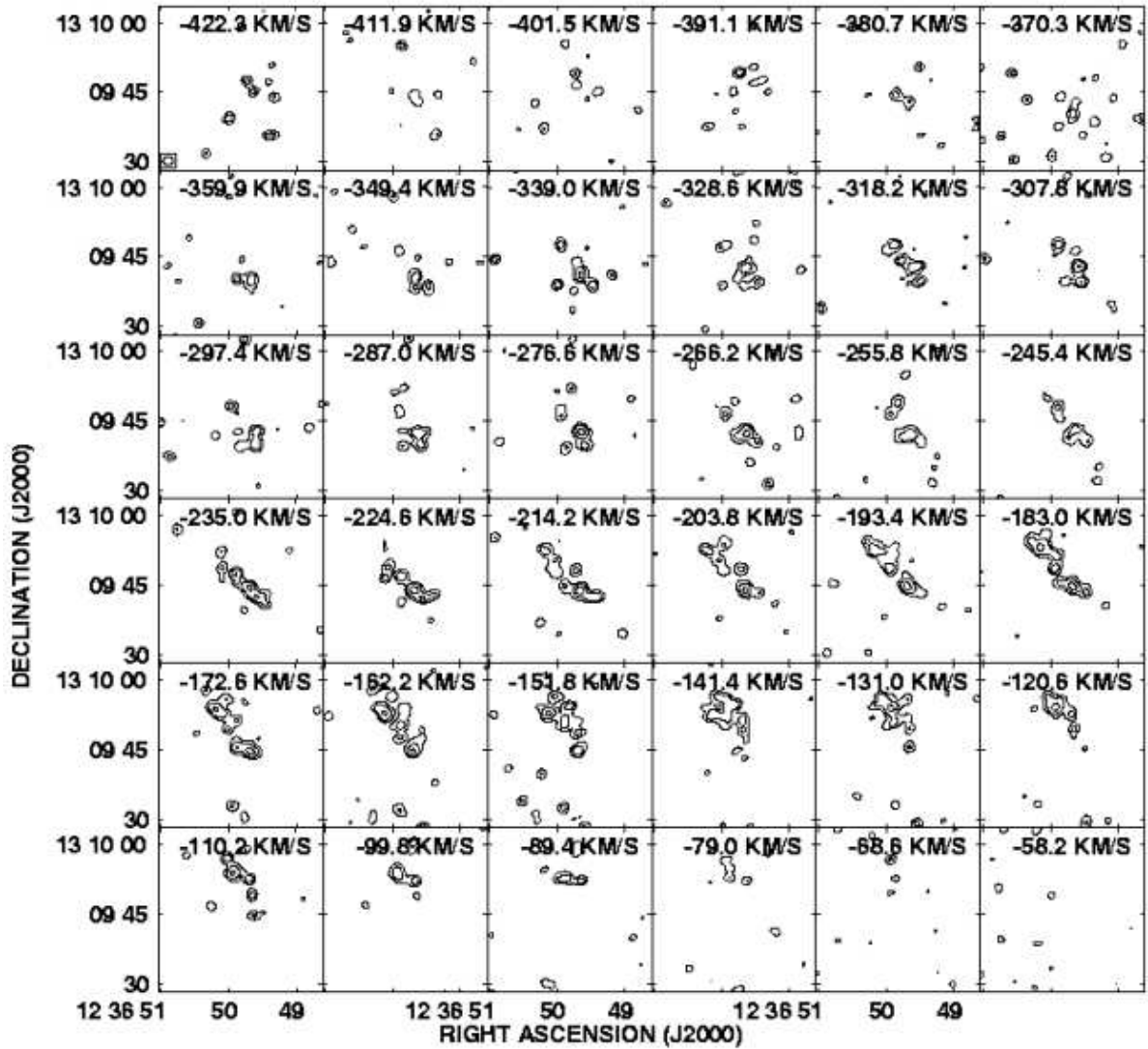}
  \end{center}
  \caption{CO channel maps of the central $35\arcsec$ obtained with uniform weighting. The first channel at the top-left corner corresponds to a radial velocity of $-422.3$ km s$^{-1}$, and the last one at the bottom right corner corresponds to $-58.2$ km s$^{-1}$. The contour levels are 1.5, 3, 6, 12 $\sigma$ ($\sigma =34$ mJy beam$^{-1}$). \label{chmap}}
\end{figure*}

Figure \ref{chmap} shows channel maps of the central $35\arcsec$ region obtained with a uniform weighting. The increment of contiguous channels is 10.4 km s$^{-1}$, although the maps were the averages over a velocity width of 20.8 km s$^{-1}$. 
The CO emissions move from south-west to north-east as the radial velocity changes from blue-shift to red-shift, which indicates rotation of the galaxy. 

\vskip 3mm
\centerline{--- Figure \ref{chmap} ---}
\vskip 3mm

\subsection{CO Intensity Maps}
\label{sec-map}
We derived integrated CO intensity maps (zeroth moment maps) from the obtained data cubes using the task {MOMNT} in {AIPS}. {We show two maps, i.e. low- and high-resolution maps, derived using the two different weighting functions in CLEAN.}

\subsubsection{Natural weighting map}
\label{sec-namap}
\begin{figure*}
  \begin{center}
    \FigureFile(80mm,160mm){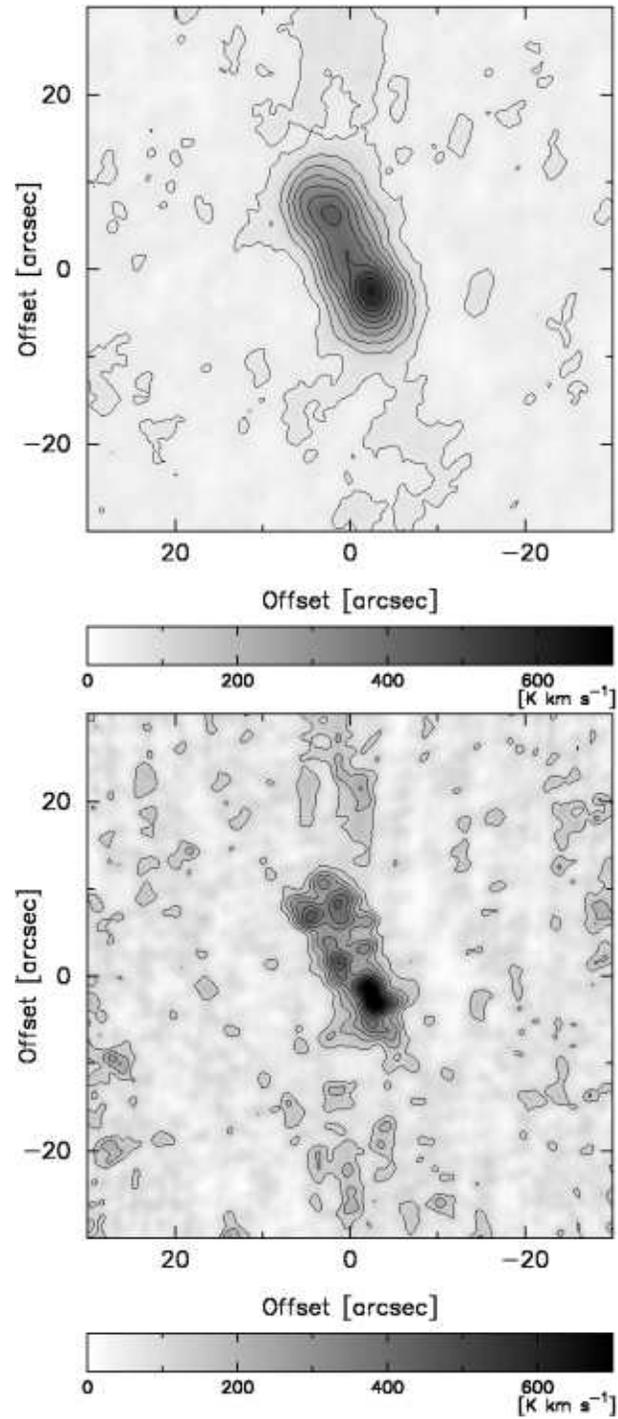}
  \end{center}
  \caption{CO velocity-integrated intensity map. Left panel: CO map obtained with natural weighting. The contour levels are 60, 120, 180, 240, 300, 360, 420, 480, 540, and 600 K km s$^{-1}$. Right panel: CO map obtained with uniform weighting. The contour levels are 120, 180, 240, 300, 360, 420, 480, 540, and 600 K km s$^{-1}$. \label{MOM0}}
\end{figure*}

The low-resolution map, {which was obtained using a natural weighing function}, is shown in the top panel of figure \ref{MOM0}. 
The molecular gas is highly concentrated into the central compact region. This molecular disk is elongated in the north-east (south-west) direction. 
The semi-major axis radius is about $12\arcsec$ by tracing the contour of 120 K km s$^{-1}$ in the top panel of figure \ref{MOM0}, which corresponds to 1.6 kpc. The molecular disk has two off-center peaks, and the south-western one is brighter than the other. {In addition, two diffuse ridges run from the central concentration toward the north and south directions. These are the so-called offset ridges, which are often seen in barred galaxies. } 

\vskip 3mm
\centerline{--- Figure \ref{MOM0} ---}
\vskip 3mm

\begin{figure*}
  \begin{center}
    \FigureFile(80mm,80mm){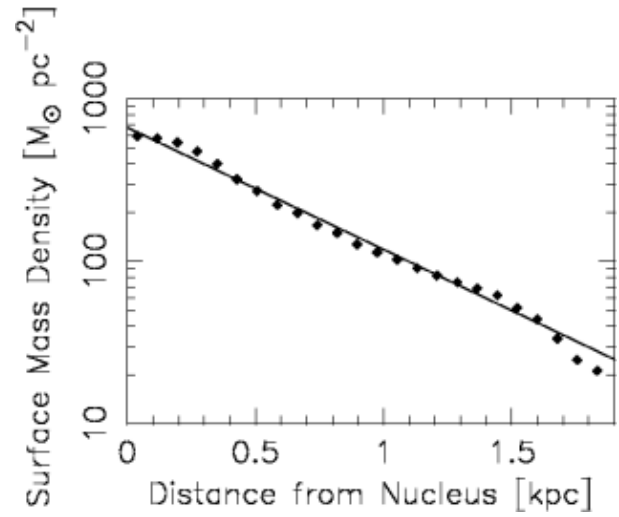}
  \end{center}
  \caption{Radial profile of the H$_2$ molecular gas distribution. The vertical axis is logarithmic scale. The fitted line denotes $\Sigma [M_\odot {\rm pc}^{-2}] = 670 \exp{(-R/0.58 ({\rm kpc}))}$.  \label{radpro}}
\end{figure*}

We present a radial distribution of the molecular gas in figure \ref{radpro}, which was obtained by using the task {IRING}. To make this figure, the primary beam attenuation was corrected. The surface density was calculated by adopting a CO-to-H$_2$ conversion factor of $C = 2.1 \times 10^{20}$ cm$^{-2}$ K$^{-1}$ km s$^{-1}$ (Arimoto et al. 1996) and multiplying by $\cos{63\arcdeg}$ to take into account the inclination. The radial distribution of the molecular surface mass density, $\Sigma$ [$M_\odot$ pc$^{-2}$], can be well reproduced with an exponential function,
\begin{equation}
\label{eq-exp}
\Sigma = \Sigma_0 \exp{(-R/R_e)},
\end{equation}
where $\Sigma_0$ is the central surface density ($670$ $M_\odot$ pc$^{-2}$), $R$ is the radius, and $R_e$ is the $e$-folding radius of 0.58 kpc. {Integrating this function,} the total gas mass, i.e. within the radius of $12\arcsec=0.94$ kpc, is $6.8\times 10^8$ $M_\odot$. The data do not suffer from the missing flux (see subsection 3.1).

The CO distribution presented in the top panel of figure \ref{MOM0} is consistent with \citet{sak99}.

\vskip 3mm
\centerline{--- Figure \ref{radpro} ---}
\vskip 3mm

\subsubsection{Uniform weighting map} 
\label{sec-uni}
\begin{figure*}
  \begin{center}
    \FigureFile(160mm,80mm){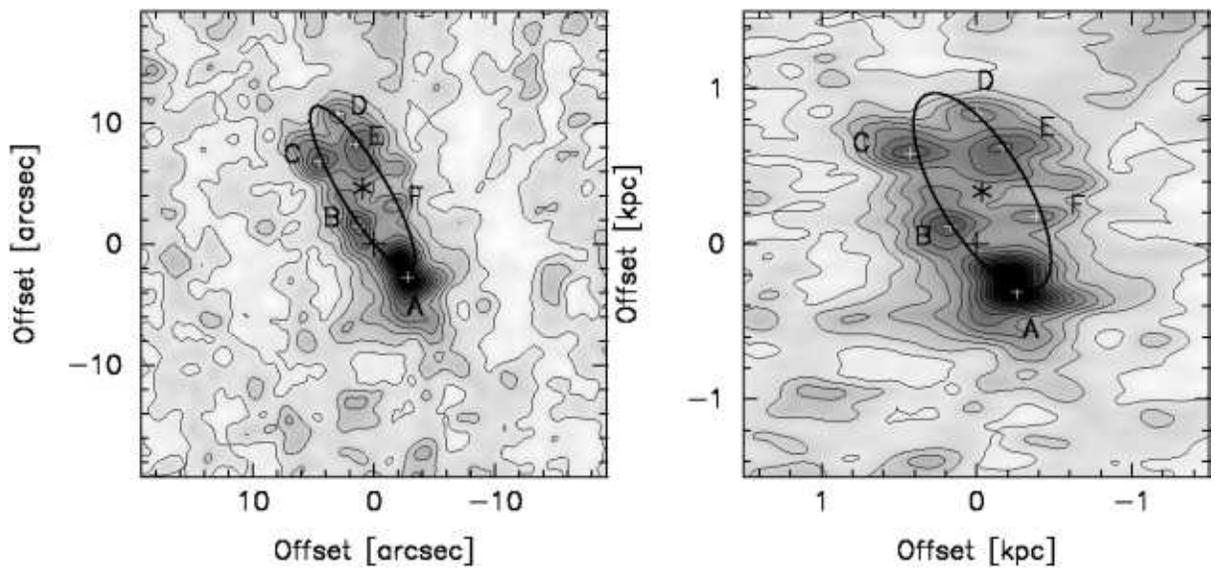}
  \end{center}
  \caption{Elliptical fitting of the CER. The thick line shows the best-fitted ellipse, whose center is indicated by asterisk symbol. The black cross shows the central position determined by 2MASS image \citep{jar03}. Six CO blobs are indicated by figures of 'A', 'B', 'C', 'D', 'E', and 'F'. Their peak positions are indicated by white crosses. Left: On the sky. Right: Face-on view. \label{blob-sp}}
\end{figure*}

The highest resolution map, {which was obtained with a uniform weighing function}, is shown in the bottom panel of figure \ref{MOM0}.
The map {shows that the molecular gas is distributed like an elliptical ring (figure \ref{blob-sp}).} 
This feature is in fact elliptical in a face-on view after correcting for the inclination using the same inclination angle and position angle of the major axis as the main disk. A face-on view of this ring is shown in {the right panel of} figure \ref{blob-sp}. We call this feature the circumnuclear elliptical ring (CER). 
CER consists of six major molecular blobs with peak intensities brighter than 300 K km s$^{-1}$. {The formations of the CER and these blobs may be related to the central activity, as suggested by Wada and Habe (1992). We discuss this idea in section \ref{sec-discussion}. } We name these blobs 'A', 'B', 'C', 'D', 'E' and 'F', which are indicated in figure \ref{blob-sp}. The peak positions of these blobs are indicated by white crosses. 
In order to calculate the masses of the blobs, we summed up the larger intensities than thresholds: 300 K km s$^{-1}$ for blobs 'D' and 'F', and 360 K km s$^{-1}$ for the others. The masses range from $5.44\times 10^6 M_\odot$ to $1.39\times 10^8 M_\odot$, which were corrected for the missing flux of $29\%$ and the primary beam attenuation. The peak-positions, peak-integrated intensity, and the masses are given in table \ref{parm-blob}. 

In order to measure the radii of the major and minor axes, the axial ratio, the central position, and the position angle of the CER, we fitted an ellipse into the CER.
The central position was fixed at the mean position of the six blobs, and then, the radii and the position angles of the major and minor axes were calculated by the least squares method. The central position of this ellipse is indicated with an asterisk symbol in figure \ref{blob-sp}, which is RA(J2000) $= 12^{\rm{h}}36^{\rm{m}}49\fs9$, Dec(J2000) $= +13\arcdeg 09\arcmin 50\arcsec$. This is offset by 340 pc from the central position, determined by 2MASS image \citep{jar03}. On the sky map the radii of the major and minor axes of the CER are $7\farcs8$ and $1\farcs9$, respectively, and the axial ratio is 0.24. The position angle of this ellipse is $31\arcdeg$ from the north. In the case of a face-on view, the radii of the major and minor axes of the CER are 720 and 280 pc, respectively, and the axial ratio is 0.39. The position angle of this ellipse is $31\arcdeg$ from the major axis of the optical disk when we look at this galaxy face-on. The parameters of the CER are given in table \ref{parm-CER}.

We compare this high-resolution map with optical broadband, H$\alpha$, and H {\sc i} data in subsection \ref{sec-comp}.
Formations of these blobs may result in collisional dissipation between blobs and central starburst activity. 
We discuss the formation of large molecular blobs and the central activity in section \ref{sec-discussion}.

\vskip 3mm
\centerline{--- Figure \ref{blob-sp} ---}
\vskip 3mm

\vskip 3mm
\centerline{--- Table \ref{parm-CER} ---}
\vskip 3mm

\subsection{Velocity Field}
\label{sec-vel}
\begin{figure*}
  \begin{center}
    \FigureFile(160mm,80mm){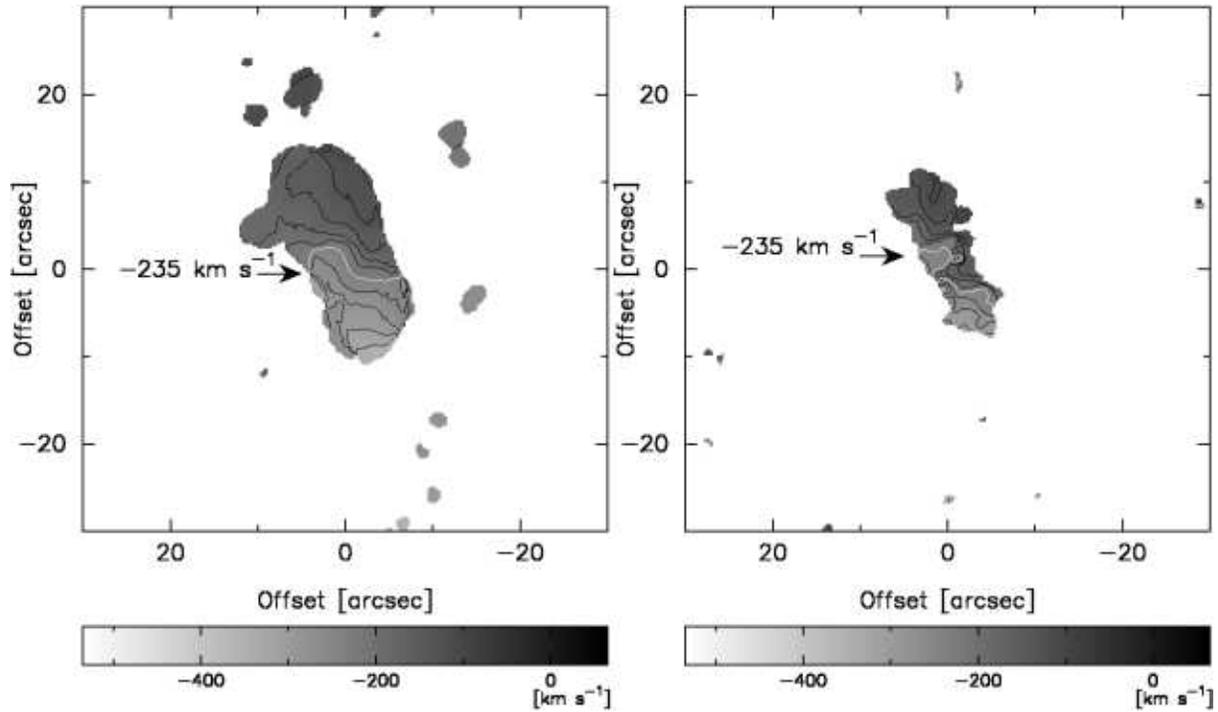}
  \end{center}
  \caption{Velocity fields obtained with natural weighting  (left) and with pure uniform weighting (right). Contour levels are $-335$, $-310$, $-285$, $-260$, $-235$, $-210$, $-185$, $-160$, and $-135$ km s$^{-1}$. The white line indicates $-235$ km s$^{-1}$.\label{MOM1}}
\end{figure*}

Figure \ref{MOM1} shows intensity-weighted velocity fields (first moment maps). The images at the top and bottom are made from low- and high-resolution cubes. The parameters used to make these images are given in table \ref{parm-vf}. The northern part of the disk is red-shifted relative to the systemic velocity ($-235$ km s$^{-1}$) and the southern part is blue-shifted. Hence, the north-western side of the disk is in the near side, assuming trailing spiral arms.

{The velocity field largely deviates from a spider diagram, i.e. pure circular rotation. The isovelocity contours show a strong $S$-shaped twist. 
The distortion of the velocity field must result from a bar-potential.}

\vskip 3mm
\centerline{--- Table \ref{parm-vf} ---}
\vskip 3mm

\vskip 3mm
\centerline{--- Figure \ref{MOM1} ---}
\vskip 3mm

\subsection{PV Diagram}
\label{sec-pv}
\begin{figure*}
  \begin{center}
    \FigureFile(100mm,240mm){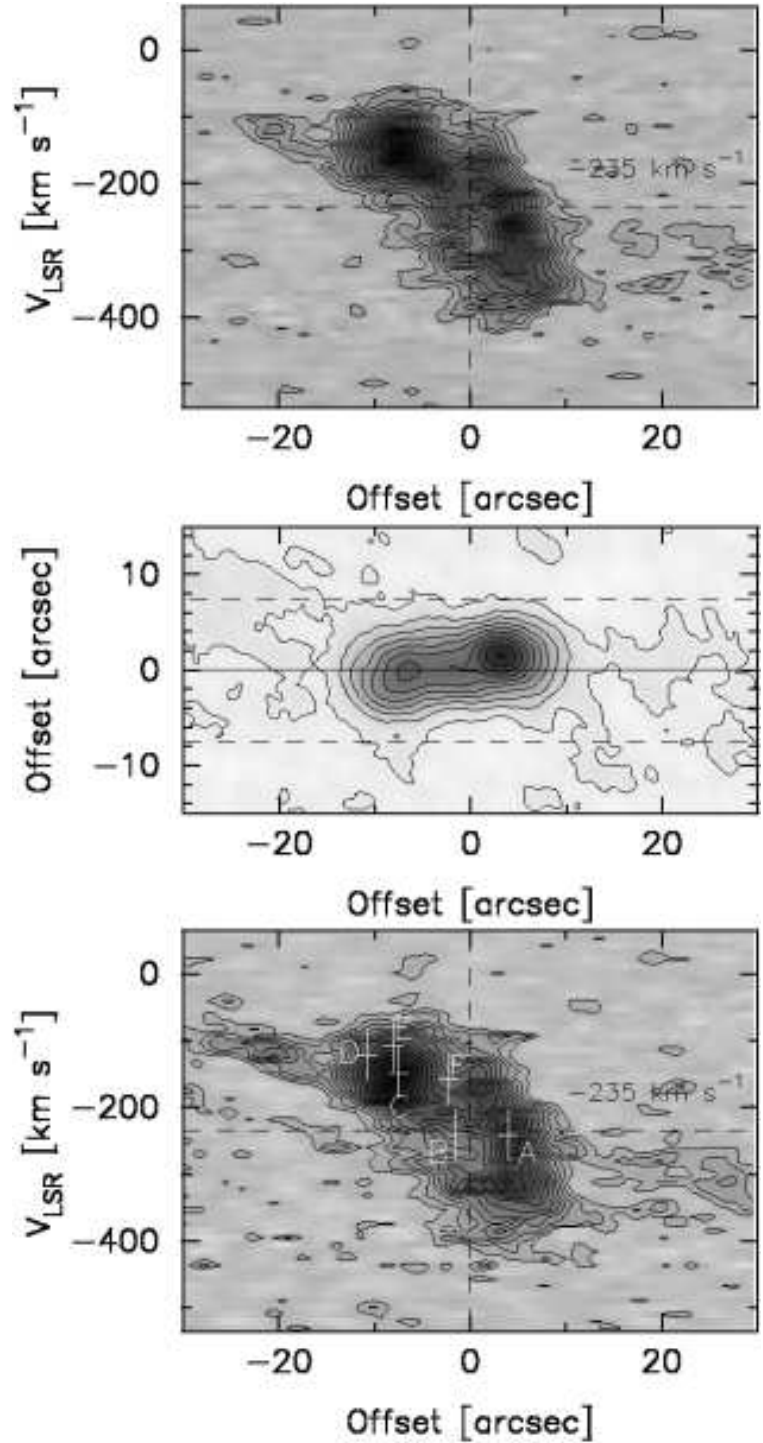}
  \end{center}
\caption{PV diagrams along the major axis made from the low-resolution cube. The central panel is a low-resolution CO image, which was rotated by $72\arcdeg$ counter-clockwise so that its major axis is horizontal. The top panel was obtained by slicing the cube along the major axis, which is shown as the central thick line in the central panel. The bottom panel shows an averaged PV diagram over $15\arcsec$ width, which is shown by two dashed lines in the central panel. \label{PV}}  
\end{figure*}

PV diagrams along the major axis at a position angle of $18^\circ$ are shown in figure \ref{PV}. 
{We adopted RA(J2000) $= 12^{\rm{h}}36^{\rm{m}}49\fs8$, Dec(J2000) $= +13\arcdeg 09\arcmin 46\farcs3$ as the central position, which was determined using the 2MASS $K$-band image \citep{jar03}.} 
The central panel of figure \ref{PV} is a low-resolution CO image, which was rotated by $72\arcdeg$ counter-clockwise so that its major axis is horizontal. The top panel of figure \ref{PV} was obtained by slicing the cube along the major axis (thick line in the central panel). The bottom panel of figure \ref{PV} shows a PV diagram sliced with the $15\arcsec$ width (two dashed lines). Significant forbidden velocity components were found in the PV diagrams, i.e. the blue-shifted component ($V_{\rm lsr}<-235$ km s$^{-1}$) at the an offset $< 0\arcsec$ and a red-shifted component ($V_{\rm lsr}>-235$ km s$^{-1}$) at an offset $> 0\arcsec$. These forbidden velocities indicate the presence of non-circular motion. Both PV diagrams are similar to a parallelogram. 

The bottom panel shows the positions and velocities of blobs. The vertical and horizontal bars denote the velocity (
76 km s$^{-1}$) and spatial resolutions ($2\arcsec$), respectively. {The velocity of each blob is measured using the bottom panel of figure \ref{MOM1}. These values are given in table \ref{parm-blob}. The blobs lie along a parallelogram, which implies that the blobs do not circularly rotate.}

\vskip 3mm
\centerline{--- Figure \ref{PV} ---}
\vskip 3mm

\vskip 3mm
\centerline{--- Table \ref{parm-blob} ---}
\vskip 3mm

\begin{figure*}
    \begin{flushleft}
    \FigureFile(160mm,160mm){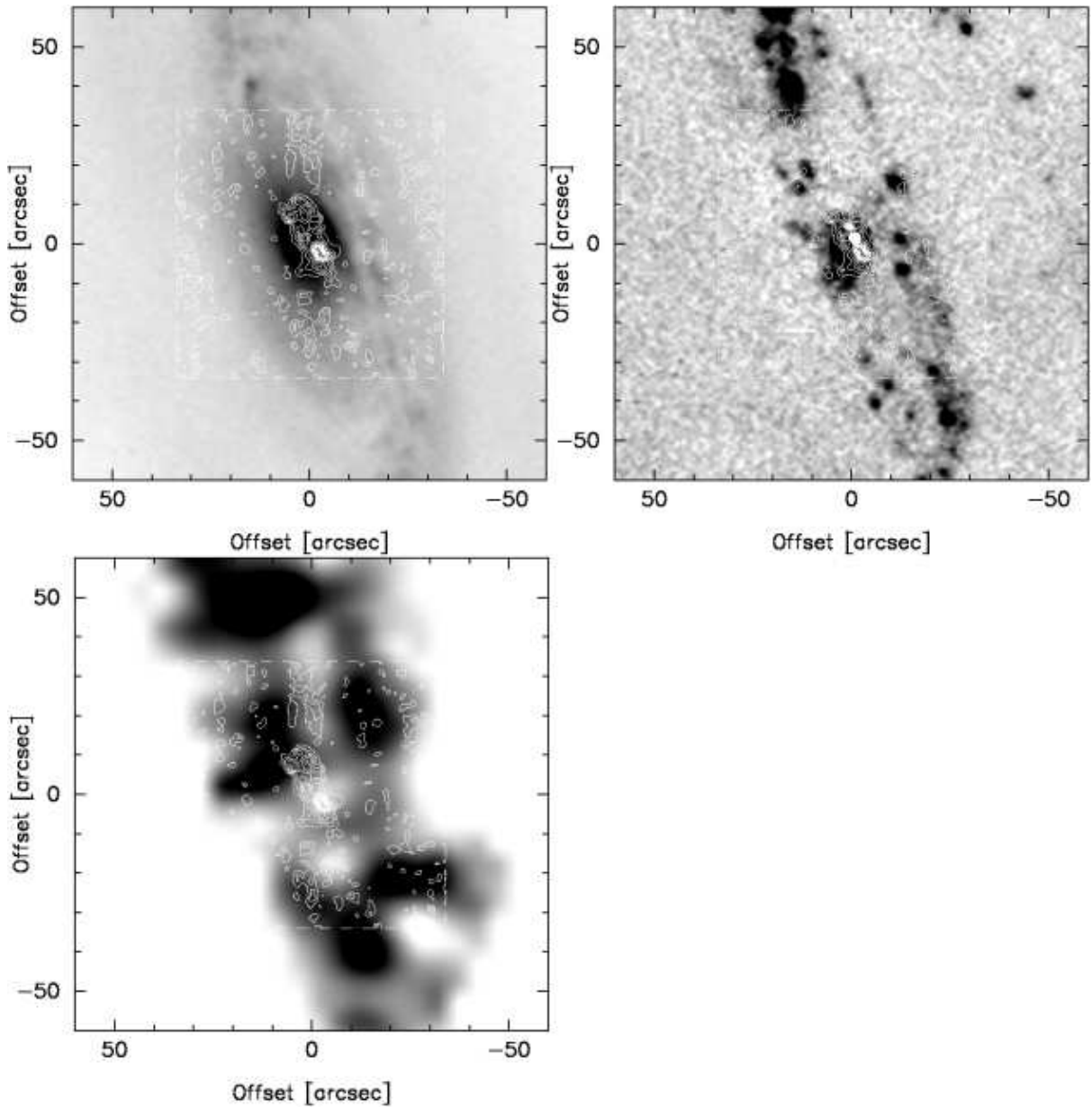}
    \end{flushleft}
  \caption{CO image with uniform weighting superimposed on images with other wavelengths. The square box with the dashed line denote the central $1\arcmin \times 1\arcmin$ in each box. Top-left: CO map (contour) and $R$-band image (gray), which is taken from \citet{koo01}.  
Top-right: CO map (contour) and H$\alpha$ image (gray), which is taken from \citet{koo01}. Bottom-left: CO map (contour) and H {\sc i} image (gray), which is taken from \citet{cay90}. \label{comparison}}
\end{figure*}

\begin{figure*}
  \begin{center}
    \FigureFile(160mm,80mm){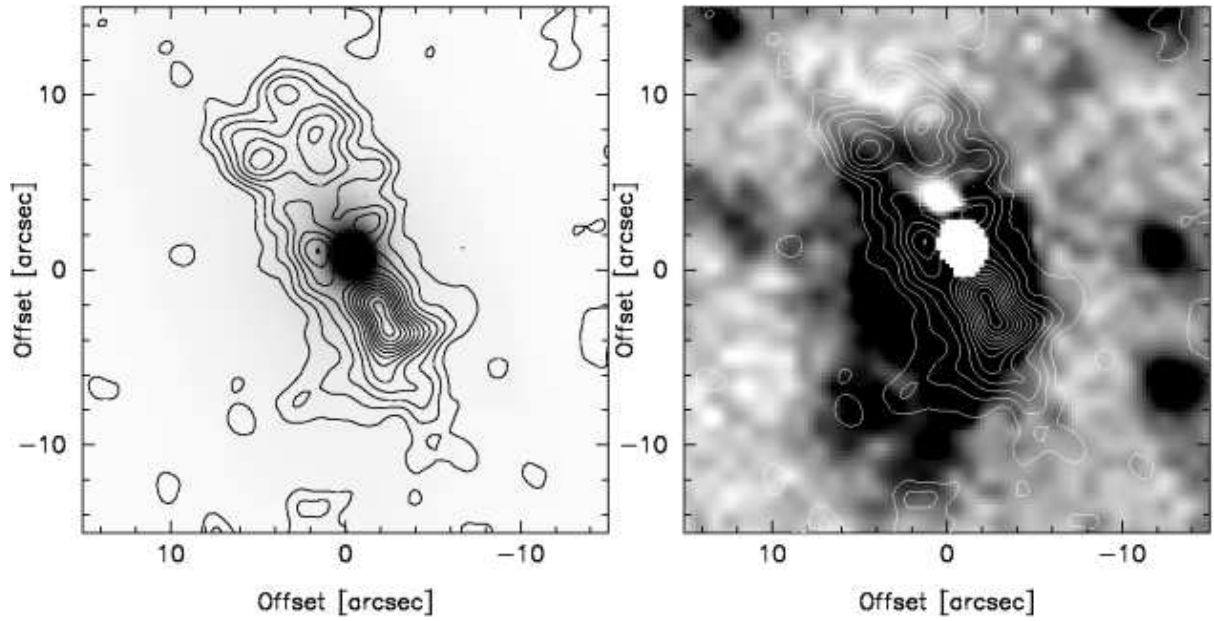}
  \end{center}
  \caption{Magnified images of the central regions of figure \ref{comparison}. Left panel: The central region of the CO (contour) and $R$-band image (gray), which is the top-left panel of figure \ref{comparison}, but with a different contrast. Right panel: The central region of the CO (contour) and H$\alpha$-band image (gray), which is the top-right panel of figure \ref{comparison}. \label{comparison2}}
\end{figure*}

{
\subsection{Comparisons with Other Observations}
\label{sec-comp}
\subsubsection{$R$-band image}
The top-left panel of figure \ref{comparison} shows the CO image superimposed on an $R$-band image. The $R$-band and H$\alpha$ images are taken from \citet{koo01}. The absolute astrometry of these images was determined by comparing them with the DSS $R$-band image. 

The left panel of figure \ref{comparison2} shows the central region of this $R$-band image, but with a different contrast. The optical bright nucleus is located inside the molecular CER.

\subsubsection{H$\alpha$ image}
The top-right panel shows the CO image superimposed on the H$\alpha$ image whose astrometry was adjusted with the above-mentioned $R$-band image \citep{koo01}. The H$\alpha$ emission consists mainly of the central bright core and a ring with a radius of $\sim 1\arcmin$. The ring is not axisymmetric about the center. 
The H$\alpha$ ring is shifted toward the north-west direction. The central bright core is located at the eastern part of the H$\alpha$ ring. The right panel of figure \ref{comparison2} shows the central region of this image. The CER is located at the bright H$\alpha$ core. The central H$\alpha$ bright core implies that a central starburst is occurring. We discuss the relationship between the bright H$\alpha$ core and the CER in the next section. 

\subsubsection{H {\sc i} Image}
The bottom-left panel of figure \ref{comparison} shows the CO image superimposed on the H {\sc i} image taken from \citet{cay90}. Note that the spatial resolution of the H {\sc i} data ($17\arcsec \times 13\arcsec$) is much larger than the resolution of our CO data ($2\farcs0 \times 1\farcs8$). The H {\sc i} gas forms a ring-like structure, which is similar to the H$\alpha$ image. The molecular CER is located at the central-east part of the H {\sc i} ring. 

}

\vskip 3mm
\centerline{--- Figure \ref{comparison} ---}
\vskip 3mm

\vskip 3mm
\centerline{--- Figure \ref{comparison2} ---}
\vskip 3mm

\section{Discussion}
\label{sec-discussion}
The highest resolution CO map {shows that six molecular blobs are likely to be the CER.} There is a bright H$\alpha$ core at the center of the CER. The existence of the CER is common and is sometimes found in spiral galaxies (e.g., Kohno et al., 2003). The formation of the CER may be related to the dynamical condition, which we discuss in the following subsection. Moreover, we discuss the existence of molecular blobs and the bright H$\alpha$ core accompanying the CER. 

\subsection{Formation of the CER}
The velocity field (figure \ref{MOM1}) and the PV diagram (figure \ref{PV}) show the existence of a strong non-circular motion. \citet{nak04} showed that the two-dimensional orbits of the ISM of this galaxy are in fact oval. This non-circular motion implies that there is an asymmetric potential (bar structure). Although the optical image of NGC 4569 does not show a strong bar, the non-circular motion can be reproduced even by a weak-bar potential  (e.g., Wada, Habe, 1992; Koda, Wada, 2002). 
The central gas dynamics is largely related to the radii of the inner Lindlbad resonances (ILRs). Wada and Habe (1992) have shown that the gas forms a CER around ILRs. NGC 4569 might have ILRs and the CER has been formed as a result.

\begin{figure*}
  \begin{center}
    \FigureFile(160mm,240mm){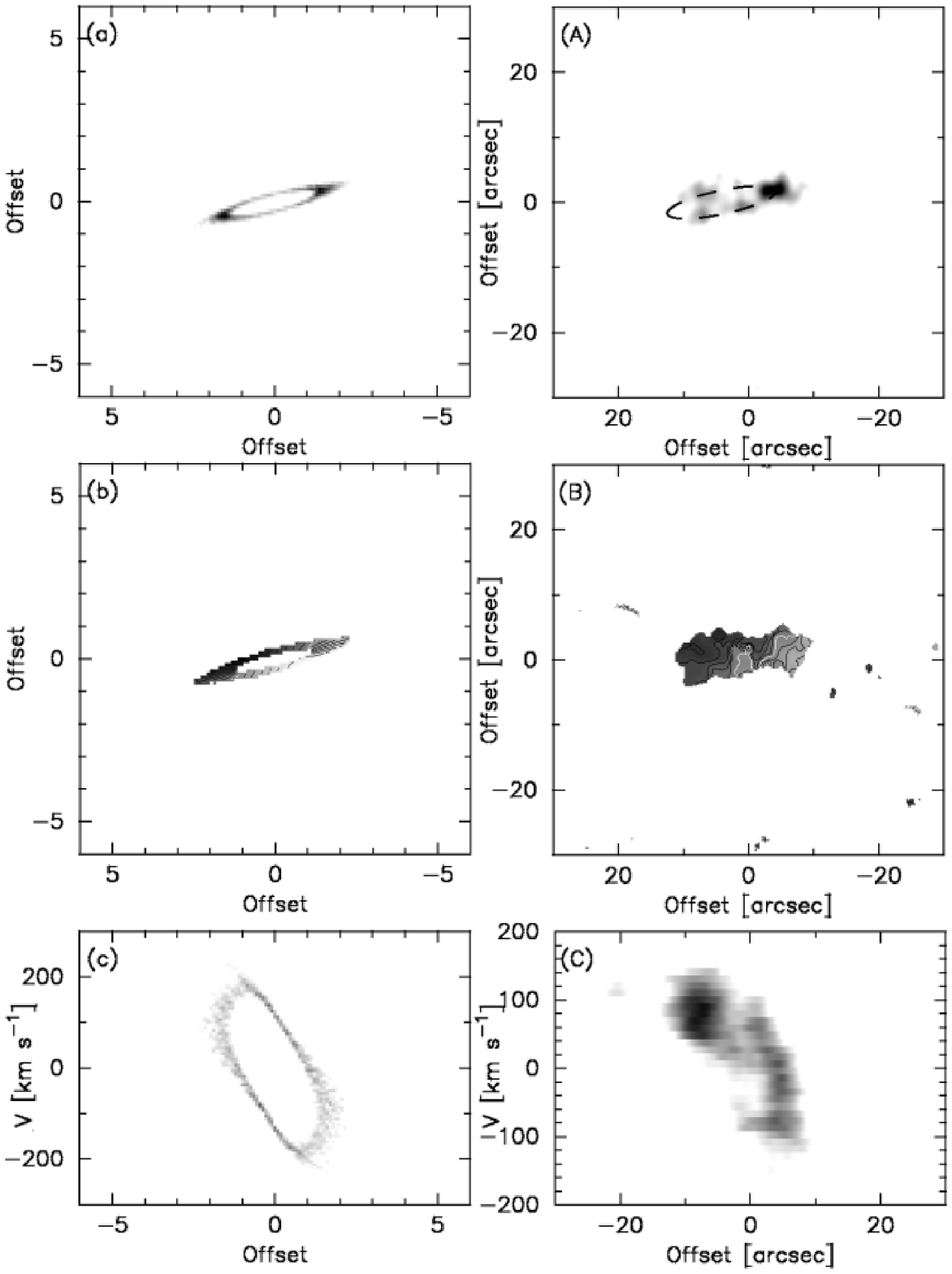}
  \end{center}
\caption{Comparisons of the SPH simulation presented by Koda and Wada (2002) and the observed data. Left-hand-side images are results of an SPH simulation; (a) Gas distribution, (b) Velocity field, and (c)  Position--velocity diagram. Right-hand-side images are observational data; (A) High resolution CO image rotated by $72\arcdeg$ counterclockwise, (B) Low-resolution velocity field rotated like (A), and (C) Position--velocity diagram. \label{OBS-SIM}}
\end{figure*}

{We compare the observational data with a Smoothed Particle Hydrodynamics (SPH) simulation of a gas disk in a bar potential. Figures \ref{OBS-SIM} a, b, and c show snapshots of the gas distribution, and velocity field, and a PV diagram from model E of Koda and Wada (2002), respectively. We rotated the gas distribution and the velocity field so that their major axes are aligned with the horizontal axes of figures \ref{OBS-SIM}a and b using a position angle of $18\arcdeg$ and an inclination of $63\arcdeg$. The major axis of the elliptical ring is assumed to be tilted counterclockwise by $31\arcdeg$ on the plane of the galaxy disk (see subsubsection \ref{sec-uni}). 
There is a strong concentration of the gas on a central elliptical ring, and the gas is moving along the ring (figure \ref{OBS-SIM}a).

Figure \ref{OBS-SIM}b shows that the global velocity gradient is not parallel to the minor axis of the main disk, which implies that the gas does not circularly rotate.
A PV diagram of the CER is shown in figure \ref{OBS-SIM}c, which shows forbidden velocity components as blue-shifted components in the left-hand side and red-shifted components in the right-hand side. The PV diagram is similar to a parallelogram. As shown in figure \ref{PV}, the gas in the CER moves along this parallelogram. 

Corresponding figures are presented in the right-hand side of figure \ref{OBS-SIM}; figures \ref{OBS-SIM}A, B, and C are the gas distribution, velocity field, and PV diagram obtained from observations, respectively.

The observed features for NGC 4569 are well reproduced by the SPH simulation; the observed ring structure in the higher resolution map (figure \ref{OBS-SIM}A) is similar to the elliptical ring of figure \ref{OBS-SIM}a. The $S$-shaped twists in the observed velocity field (figure \ref{OBS-SIM}B) are similar to that of the model (figure \ref{OBS-SIM}b). 
The parallelogram-like feature seen in the observed PV diagram (figure \ref{OBS-SIM}C) is also reproduced by the model (figure \ref{OBS-SIM}c).}

In the next subsection, we discuss the gaseous self-gravity and the formation of large molecular blobs. 

\vskip 3mm
\centerline{--- Figure \ref{OBS-SIM} ---}
\vskip 3mm

\subsection{Large Molecular Blobs and Central Starburst}
 
Wada and Habe (1992) discussed that the CER fragments and collapses due to the self-gravitational instability of gas when the ratio of the gas mass to the stellar mass is larger than 10\%. They described the process of decay of the CER as follows: (1) If the CER is dense enough, it fragments and collapses into blobs due to the self-gravity. 
(2) Then, these gas blobs collide with each other. 
Kinetic energy is dissipated due to the collisions. (3) Massive blobs are formed by accumulating smaller blobs, losing the kinetic energies. (4) Consequently, the large blobs fall into the center. 

The observed results are likely to support this scenario. First, the CER and non-circular motion are consistent with the gas distribution and velocity field caused under a weak bar potential in numerical hydrodynamic simulations. Second, the large molecular blobs observed in this galaxy are also found in simulations. 
The largest blob A is the closest to the galactic center, which can be understood by the scenario: small blobs grow into a larger one while losing the kinetic energies due to collisions; then, the larger blob falls into the central region because of the loss of kinetic energy. 
Wada and Habe (1992) discussed that a gravitational instability occurs when the gas-to-stellar mass ratio within a radius is greater than 10\%. This condition is 
satisfied in this galaxy as follows. 
The rotational velocity is $\sim 130$ km s$^{-1}$ at the radius of $20\arcsec$, adopting an inclination angle of $63\arcdeg$ (see figure \ref{PV}). The dynamical mass within $20\arcsec$ is calculated to be $6.1 \times 10^{9} M_\odot$. Using the equation (\ref{eq-exp}), the gaseous mass within $20\arcsec$ was $1.1 \times 10^{9} M_\odot$. The gas-to-stellar mass ratio reaches $18\%$. Therefore, the self-gravity of the gas is important and a self-gravitational instability can occur.

Thus, the observable results support the scenario that the gas becomes clumpy due to its self-gravity, accumulated into the center, and is leading to the central starburst (Wada, Habe, 1992). 
The central bright H$\alpha$ core might be caused by the gravitational instability of the CER and consequent gas inflow.

\section{Summary}
We observed the central region of NGC 4569 in the $^{12}$CO ($J=1$--$0$) line using NMA. The high-resolution ($1\farcs8$ -- $2\farcs0$) CO image showed that 
the six blobs form the CER with a radius of $720$ pc. 
The velocity field and PV diagram of the CER show the presence of strong non-circular motions, which may result from a central weak bar. 
Compared with the H$\alpha$ image, we found that the CER is located in the H$\alpha$ bright central core. The gas-to-stellar mass ratio within a radius of $20\arcsec$ is $18\%$. 

The results, i.e. (1) the blobs within the CER, (2) non-circular motions, (3) the coincidence of the CER and the bright H$\alpha$ core, and (4) the large gas-to-stellar mass fraction, support the scenario of gaseous inflow driven by a bar potential and the gaseous self-gravity (Wada, Habe, 1992). 

\vskip 5mm

We are grateful to the staff members of the NRO for their helpful discussions about the observation and reduction. We thank Dr. K. Kohno, Dr. S. K. Okumura, Dr. Y. Tutui, Mr. T. Takamiya, Mr. M. Hidaka, and Ms. S. Onodera for their cooperation in the observations. We are grateful to Dr. V. Cayatte for providing us with their H {\sc i} data. We thank Dr. R. Koopmann for providing us with the $R$- and H$\alpha$ data. 


\newpage

\begin{table}
\begin{center}
\caption{Properties of NGC 4569. \label{prop-n4569}}
\vskip 2mm
\begin{tabular}{lcc}
\hline\hline
Parameter & Value & Reference$^*$\\
\hline
Morphology & SAB(rs)ab & 1\\
Position    &   & 2\\
 \hskip 5mm RA(J2000) & $ 12^{\rm h}36^{\rm m}49\fs8$&\\ 
 \hskip 5mm Dec(J2000) & $ +13\arcdeg09\arcmin46\farcs3$&\\
Total {\it B} Magnitude& 10.26  mag & 1\\
Systemic velocity& $-235$ km s${}^{-1}$& 1\\
Assumed distance& 16.1 Mpc & 3\\
  & $1'' = 78$ pc  & \\
Inclination angle & $63\arcdeg$ & 4\\
Position angle & $18\arcdeg$&4\\
\hline
\end{tabular}\\
$^*$(1) RC3 \citep{vau91}; (2) \citet{jar03}; (3) \citet{fer96}; (4) \citet{nak04}.  
\end{center}
\end{table}

\begin{table}
\begin{center}
\caption{Observation Parameters. \label{obs-parm}}
\begin{tabular}{lc}
\hline
Observed center frequency (GHz) & 115.286574  \\
Array configurations & AB, C, \& D  \\
Observing field center &    \\
 \hskip 5mm RA(J2000) & $12^{\rm{h}}36^{\rm{m}}49\fs82$\\
 \hskip 5mm Dec(J2000) & $+13\arcdeg09\arcmin45\farcs8$\\
Frequency channels & 256  \\
Total bandwidth (MHz) & 512  \\
Velocity coverage (km s$^{-1}$) & 1382  \\
Velocity resolution (km s$^{-1}$) & 10.4  \\
Amplitude and phase calibrator & 3C 273  \\
Primary beam ($\arcsec$) & 65  \\
\hline
\end{tabular}
\end{center}
\end{table}

\begin{table*}
\begin{center}
\caption{Parameters of Maps. \label{parm-map}}
\begin{tabular}{ccccccccc}
\hline\hline
Weighting  &  \multicolumn{2}{c}{Beam} & &\multicolumn{2}{c}{Velocity} & r.m.s. noise & $T_{\rm b}$ for \\
\cline{2-3} \cline{5-6}
function &  FWHM & P.A.              & & Resolution & Sampling       &$\sigma$ & 1 $\rm Jy\,beam^{-1}$\\
          &($\arcsec$)& ($\arcdeg$)            & & ($\rm km\,s^{-1}$)          &($\rm km\,s^{-1}$)  & ($\rm mJy\,beam^{-1}$)    & (K) \\
\hline
natural   & $4.5 \times 3.1$  & 146  &&  10.4 & 10.4 &  23 &  6.59\\
uniform   & $2.0\times 1.8$   & 104  &&  20.8 & 10.4 &  34 & 25.5\\
\hline
\end{tabular}\\
\end{center}
\end{table*}

\begin{table*}
\begin{center}
\caption{Parameters of CER. \label{parm-CER}}
\begin{tabular}{cccccccl}
\hline\hline
             & \multicolumn{2}{c}{Central position}                               &&\multicolumn{2}{c}{Radius}&Axial ratio & Position angle\\
\cline{2-3} \cline{5-6}
           & RA(J2000) & Dec(J2000)                                            && major & minor            &            &               \\
\hline

On the sky   & {$12^{\rm{h}}36^{\rm{m}}49\fs9$} & {$+13\arcdeg09\arcmin50\farcs5$}&& $7\farcs8$ & $1\farcs9$& 0.24& $31\arcdeg$(from north on the sky)\\
Face-on view & {$12^{\rm{h}}36^{\rm{m}}49\fs9$} & {$+13\arcdeg09\arcmin50\farcs5$}&& 720 pc& 280 pc& 0.39 & $31\arcdeg$(from major axis face-on)\\
\hline
\end{tabular}\\
\end{center}
\end{table*}

\begin{table*}
\begin{center}
\caption{Parameters of velocity fields. \label{parm-vf}}
\begin{tabular}{ccccccc}
\hline\hline
Resolution  &  \multicolumn{2}{c}{Smoothing width} & &\multicolumn{2}{c}{Velocity range} & Threshold \\
\cline{2-3} \cline{5-6}
            &  Velocity     & Spatial              & &Minimum & Maximum                     &\\
            & (km s$^{-1}$) & ($\arcsec$)             & &($\rm km\,s^{-1}$)&($\rm km\,s^{-1}$) & (K) \\
\hline
low         & 10.4          & 1.25                 & & $-437.8$ & $-42.6$ & 0.76\\
high        & 72.8          & 0.25                 & & $-422.3$ & $-58.2$ & 0.87\\
\hline
\end{tabular}\\
\end{center}
\end{table*}

\begin{table*}
\begin{center}
\caption{Parameters of blobs. \label{parm-blob}}
\begin{tabular}{cccccc}
\hline\hline
Name &  \multicolumn{2}{c}{Peak Position} & Peak intensity & Mass$^\dagger$             & Velocity\\
     &  RA(J2000) & Dec(J2000)  & (K km s$^{-1}$)  & ($M_\odot$)        & (km s$^{-1}$)\\
\hline
A    &  12$^{\rm h}$ 36$^{\rm m}$ 49$\fs$6   & 13$\arcdeg$ 09$\arcmin$ 42$\farcs$9    & 805            & $1.39\times10^8$(*1) & $-242$\\
B    &  12$^{\rm h}$ 36$^{\rm m}$ 49$\fs$9   & 13$\arcdeg$ 09$\arcmin$ 47$\farcs$4    & 481            & $2.82\times10^7$(*1) & $-242$\\
C    &  12$^{\rm h}$ 36$^{\rm m}$ 50$\fs$1   & 13$\arcdeg$ 09$\arcmin$ 52$\farcs$6    & 459            & $2.29\times10^7$(*1) & $-149$\\
D    &  12$^{\rm h}$ 36$^{\rm m}$ 50$\fs$1   & 13$\arcdeg$ 09$\arcmin$ 52$\farcs$6    & 317            & $5.44\times10^6$(*2) & $-122$\\
E    &  12$^{\rm h}$ 36$^{\rm m}$ 49$\fs$9   & 13$\arcdeg$ 09$\arcmin$ 54$\farcs$2    & 432            & $3.40\times10^7$(*1) & $-108$\\
F    &  12$^{\rm h}$ 36$^{\rm m}$ 49$\fs$7   & 13$\arcdeg$ 09$\arcmin$ 49$\farcs$1    & 335            & $6.03\times10^6$(*2) & $-158$\\
\hline
\end{tabular}\\
$^\dagger$ Threshold intensity is taken to be 360 (*1) and 300 (*2) K km s$^{-1}$.\\
\end{center}
\end{table*}

\end{document}